\documentclass[12pt,a4paper,onecolumn]{article}

\usepackage[top=2cm,bottom=2cm,left=2cm,right=2cm]{geometry} 
\usepackage{enumerate}
\usepackage{graphicx}
\usepackage{caption}
\usepackage{xcolor}
\usepackage[normalem]{ulem}

\title{Cotranslational folding of {deeply} knotted proteins}

\author{Mateusz Chwastyk and Marek Cieplak\\
\small Institute of Physics, Polish Academy of Sciences, Al. Lotnik\'ow 32/46, 02-668 Warsaw, Poland}

\begin{document}

\maketitle

\abstract{Proper folding of { deeply} knotted proteins has a very low success 
rate even in structure-based models which favor formation of the native contacts
but have no topological bias. By employing a structure-based
model, we demonstrate that cotranslational folding on a model
ribosome may enhance the odds to form trefoil knots for protein
YibK without any need to introduce any non-native contacts.
The ribosome is represented by a repulsive wall that keeps elongating
the protein. On-ribosome folding proceeds through a 
a slipknot conformation. We elucidate the mechanics and energetics
of its formation.
We show that the knotting probability in on-ribosome
folding is a function of temperature and that there is an optimal
temperature for the process. 
Our model often leads to the establishment
of the native contacts without formation of the knot.
}

\section{Introduction}

There are several hundreds of knot-containing structures
\cite{Virnau,Virnau1,Stasiak} in the Protein Data Bank (PDB) and
their topology can be characterized primarily by 
how many intersections a backbone makes with itself on making its 
two-dimensional projection. The knot extends between two end points, 
$n_1$ and $n_2 > n_1$, along the sequence. 
The knot ends are defined operationally through
systematic cutting-away of the amino acids from both termini until the
knot disintegrates \cite{km1,taylor}. The last site that still
supports the knot is its end point. If $n_1$ and $n_2$ are both distant
from the termini, the knot is considered to be deep; otherwise
it is called shallow. One needs closed lines to declare existence of
a knot on a line with certainty. Backbones of proteins are generally
not closed but for deep knots the determination of a knot-type is
quite reliable.

Stretching of a knotted protein either at constant speed
\cite{Sulkowska_2008,JACS,Dziubiella} or at constant force \cite{Israel}
results in a step-wise knot tightening process in which the knot ends jump.
Here, we consider two other conformation changing processes \cite{Dill} in
knotted proteins: folding and unfolding through heating. Folding from an
extended state to a knotted native conformation of protein YibK has been
reported \cite{dodging,Takada} to be difficult with, at best,
1 - 2\% success rate even if one uses a coarse-grained structure-based
model. Here, we examine whether nascent conditions, such that exist when
a protein is formed by the ribosome \cite{Dobson,Bustamante,ribosome,ribosome1}, 
can help in establishing the native knot in proteins. 
We propose a simple generic
model in which the ribosome is represented as an infinite repulsive plate
which spawns proteins. { Our model is an oversimplification
of the real geometry of the ribosome -- the peptide chain is formed in
the ribosome tunnel and it undergoes folding, at least partially, within it. 
The tunnel has a diameter that varies between
10 and 20 {\AA} and the largest cavity in the
tunnel cannot encompass a sphere with a radius larger than 9.5 {\AA} 
\cite{Dobson,ribosome2,ribosome3}. This geometry provides stronger
confinement than that induced by the plane. Nevertheless,
the model with the plane is the simplest one that introduces new 
qualtitative features that are brought in by confinement.}

Here, our focus is on YibK \cite{1J85} from
{\it Haemophilus influenze} which contains a deep trefoil knot with three
intersections.  
The corresponding PDB code is 1J85. (We shall refer to  
proteins through their PDB structure codes.)
{ This protein is probably the most frequently cited
example of a deeply knotted protein \cite{Virnau,Virnau1}.
Its radius of gyration is about 15 {\AA}.}
We find that the nascent conditions do help in folding 1J85
-- they actually enable it -- and the effect is observed
only within a range of temperatures that provide optimality.
We show that invoking non-native contacts is not necessary to
generate the on-ribosome slipknot, just one needs to employ a proper
procedure to define contacts that are {declared}  native.
We identify the slipknot based mechanism of folding and explain why
the model ribosome favors its formation.
We also study the energetics involved in the emergence of the slipknot.
Unlike the claims of ref. \cite{dodging} (that uses the same model),
we do not observe the slipknot in the absence of the ribosome.

\section{Structure-based modeling}

The justification and details of implementation of our model
are explained in refs. \cite{JPCM, models, PLOS}. 
Its character is
structure-based or, equivalently, Go-like \cite{Go0}, 
and the molecular dynamics deals only with the $\alpha$-C atoms.
The bonded interactions are described by the harmonic potentials.
The list of pairs of amino acids that are considered to be
in contact ({\it i.e.}  interacting)
in the native state is known as the contact map. These contacts
are described by the Lennard-Jones potentials with the minima
at the crystallographically determined distances. The potentials
are identical in depth, denoted as $\epsilon$. The value of $\epsilon$
has been calibrated by making comparisons to the experimental
data on stretching: approximately, $\epsilon$/{\AA} is 110 pN
(which also is close to the energy of the O-H-N hydrogen bond
of 1.65 kcal/mol).
Non-native contacts are considered repulsive.

The contact map itself is obtained by using the overlap criterion in which
the heavy atoms in the native conformation are represented by enlarged
van der Waals spheres \cite{JPCM,Tsai}: if at least one pair of such
spheres placed on different amino acids overlap (OV) then there is the
native contact. An alternative way to establish a contact map is
by invoking considerations that are more chemical in nature, as
available at the CSU server \cite{Sobolev}. The CSU contacts may either
be specific (like hydrogen bonds) or non-specific (presumably the much 
weaker dispersive interactions). The CSU and OV maps have many
contacts  in common but are not identical. Thus, in addition to the
OV map, we also consider OV-CSU map in which we augment CSU 
by the missing specific CSU contacts.

The backbone stiffness
is accounted for by the chirality potential \cite{JPCM}.
The simulations are done at various temperatures. For most proteins
without any knots, optimal folding takes place around
at $T=0.3~ \epsilon/k_B$,
which should correspond to a vicinity of the room temperature
($k_B$ is the Boltzmann constant).
We generate at least 200 trajectories for each temperature considered.
The time unit of the simulations, $\tau$, is effectively of order 1 ns as 
the motion of the atoms is dominated by diffusion instead of being ballistic.

Folding is usually declared when all native contacts are established
for the first time. For knotted proteins, however, this condition does 
not necessarily signify that the correct native knot has been formed.

There are two important aspect of the role of the ribosome 
in the context of nascent folding. The first is that folding
of a protein is concurrent with its birth.
Since the mRNA is translated from 5' to 3',
the proteins are synthesized from the N terminus to the C terminus.
The time interval between the emergence of two successive $\alpha$-C atoms
will be denoted by $t_w$. The second aspect is that the surface of
the ribosome provides excluded volume and, therefore, reduces the 
conformational entropy. Both aspects
can be captured by a model in which the ribosome is represented
by an infinite plate which gives birth to a protein at one fixed location.
We take the plate to generate a laterally uniform potential
of the form $\frac{3\sqrt{3}}{2}~\epsilon ~ (\frac{\sigma_0}{z})^9$,
where $z$ denotes the distance away from the plate and 
$\sigma_0 = 4\times 2^{-1/6}$ {\AA}.
This form of the potential comes from integrating the energy
of interaction between a Lennard-Jones particle 
and a semi-infinite continuum below $z$=0 and discarding the
attractive part.
A coarse-grained model of cotranslational folding with molecularly
sculpted ribosome has been proposed by Elcock \cite{Elcock}.
We have adopted a less sophisticated model in order
to enable simulations of hundreds of trajectories that last long 
-- formation of a deep knot is a rare event 
{ and making simplifications is necessary}.

\section{Results}

\begin{figure}[ht]
\begin{center}
\includegraphics[scale=0.8]{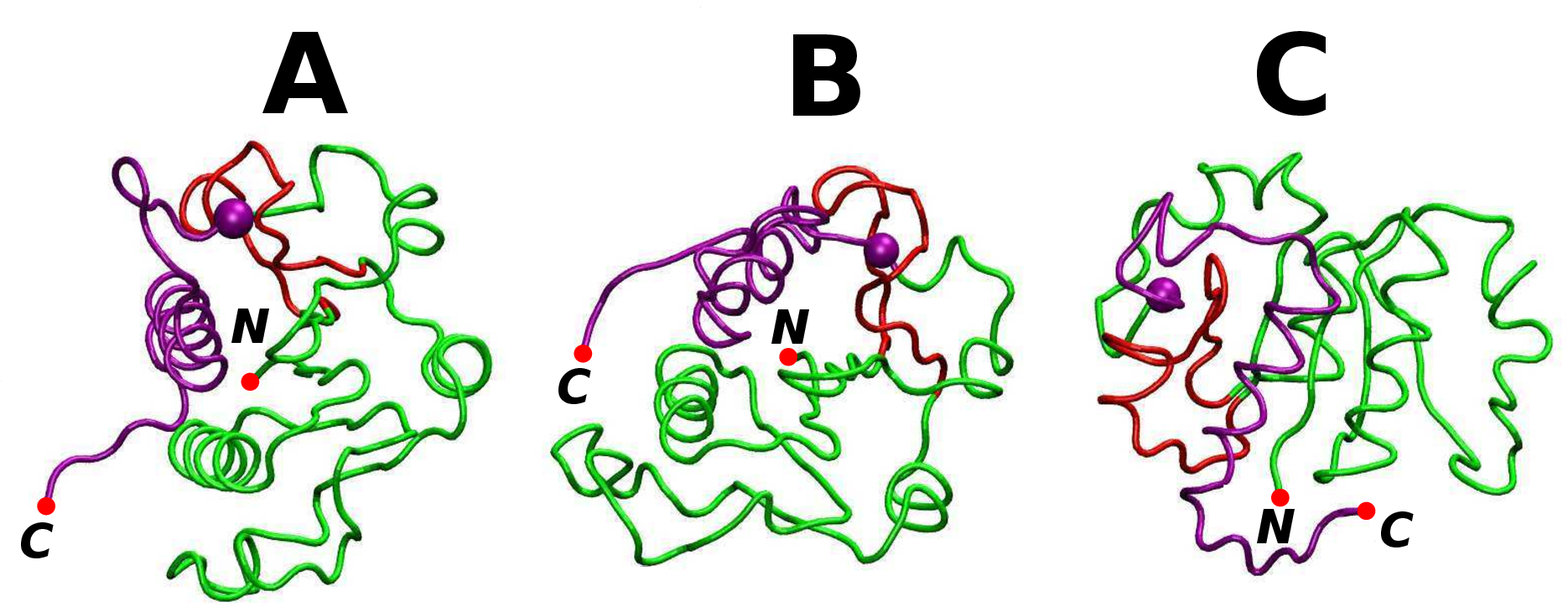}
\end{center}
  \caption{Panel A shows the native structure of protein 1J85. Panel B shows
an example of the correctly folded state obtained through cotranslational
evolution. Panel C shows an example of conformation with the 
native contacts all established but without the topology of the knot.
The meanings of the colors are explained in the main text.
}
  \label{fig:folded}
\end{figure}

The PDB structure file of 1J85 provides coordinates of 156 residues. 
About half of them are arranged into
seven $\alpha$-helices and eight $\beta$-strands \cite{1J85}.
The native conformation of 1J85 is shown in panel A of { Fig. \ref{fig:folded}}.
Segments 1-74, 75-95, 96-120,
and 121-156 are shown in green, red, green, and purple respectively.
The reason for this color convention is that in order to form
the knot in (at least) cotranslational folding, the 
purple segment (121-141)
must form a slip-loop that would go through the red knot-loop
(see the defining drawings in ref. \cite{BSDB}).
The knot ends are at LEU-75 and LYS-119 in the native state -- a separation
of 44 sites.  On stretching, the knot tightens up and the ultimate
separation between the knot ends becomes 10 \cite{Sulkowska_2008,Israel}.

\subsection{Equilibrium properties and thermal unfolding}

\begin{figure}[ht]
\begin{center}
\includegraphics[scale=0.6]{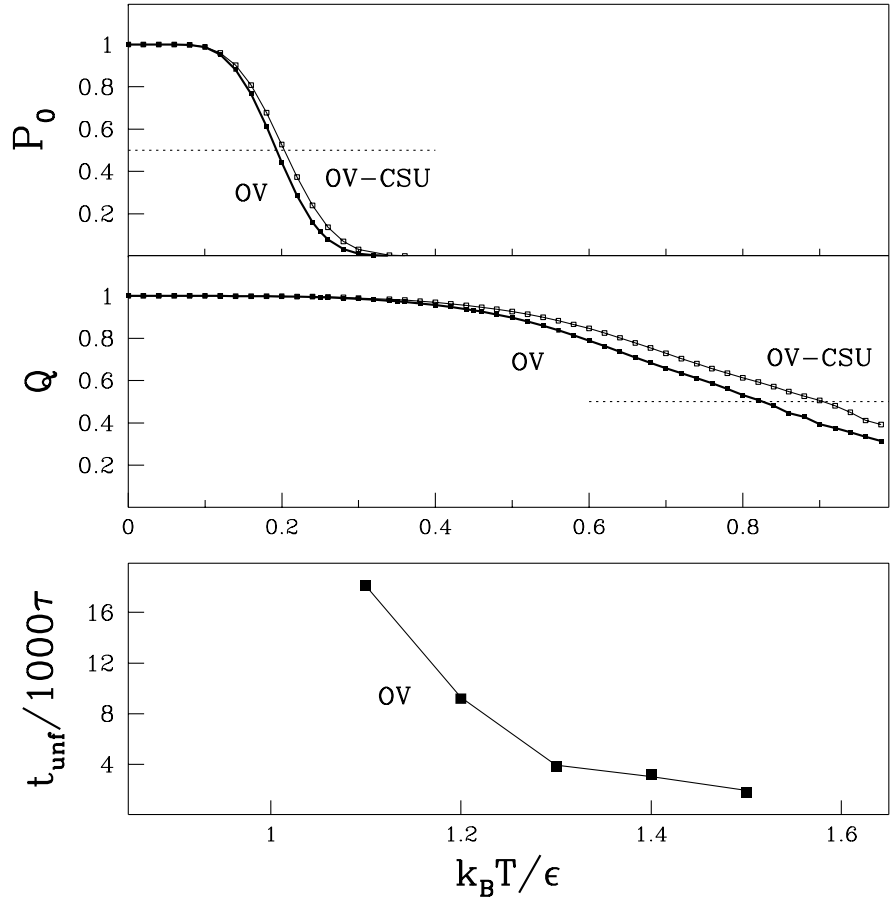}
\end{center}
  \caption{ 
The temperature-dependence of $P_0$, $Q$, and $t_{unf}$ for 1J85, 
top to bottom panels correspondingly. The contact map used is indicated.
The horizontal dotted lines in the top two panels correspond to
the value of $\frac{1}{2}$.
}
  \label{probab}
\end{figure}

In order to set the stage, we first consider a situation in which
1J85 is set in the native state and then undergoes time evolution
at various temperatures. At any finite $T$, some number of contacts
break down (the distance between the $\alpha$-C atoms in residues
that make a contact becomes larger than 1.5 of the native distance)
and some get restored. The top panel of 
Fig. \ref{probab} shows the probability, $P_0$,
of all native contacts being \emph{simultaneously} established as a 
function of $T$. Similar to the lattice models of proteins \cite{Socci}
(see also an exact analysis \cite{Banavar,Banavar1})
one may define $T_f$ as a temperature at which
$P_0$ crosses through $\frac{1}{2}$. For the OV contact map,
$T_f=0.194~\epsilon/k_B$ and for the OV-CSU contact map it is
0.204 $\epsilon/k_B$ -- just a small shift.
For both of these contact maps, $P_0$ is essentially close to zero at
$T=0.3~\epsilon/k_B$ (1\% and 3\% for OV and OV-CSU respectively). 
However, at this temperature, the fraction of the native contacts
present, $Q$, is high (the middle panel of Fig. \ref{probab}). 
$Q$ crosses $\frac{1}{2}$ at
0.827 and $0.908 ~\epsilon/k_B$ for OV and CSU-OV respectively.
Folding is said to occur when all native contacts are established
simultaneously and the kinetic optimality (typically around
0.3 - $0.35~\epsilon/k_B$) may take place where $P_0$ is small.

It should be noted that in ref. \cite{dodging} $T_f$ is defined
as corresponding to the temperature at which a half
of the native bonds are present ({\it i.e.} around 0.8~$\epsilon/k_B$)
 -- a much more relaxed
criterion compared to  $P_0$  crossing $\frac{1}{2}$. At this elevated
temperature, substantially higher than the room temperature, 
the probability of maintaing the native conformation is simply zero.
However, the temperature at which $Q$ crosses $\frac{1}{2}$
signals the onset of globular shapes in the model protein.
The simulations reported on in ref. \cite{dodging} could have
been done around the $T_f$ defined through $Q=\frac{1}{2}$ as this
is the only temperature mentioned in the text.

\begin{figure}[ht]
\begin{center}
\includegraphics[scale=0.6]{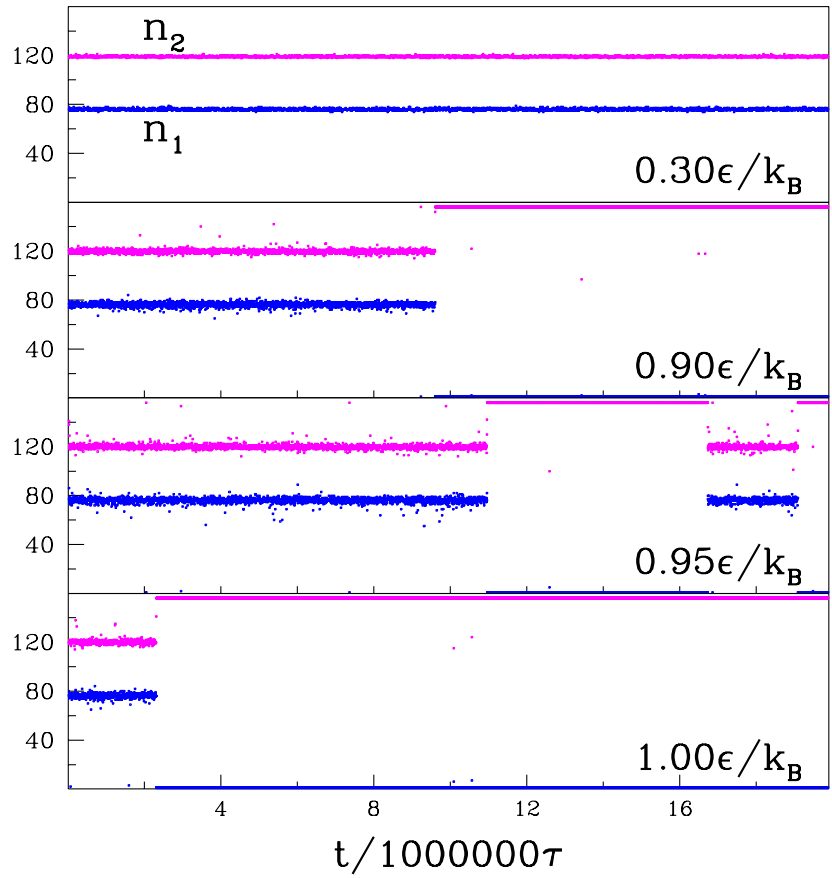}
\end{center}
  \caption{ 
Locations of the knot ends in 1J85 at various temperatures as indicated.
}
  \label{thermal}
\end{figure}

The fluctuations in the contact occupation numbers may or may not affect
the sequential locations of the knot ends. Fig. \ref{thermal} shows
locations of the knot in examples of trajectories at several 
temperatures. At $T=0.3~ \epsilon/k_B$, the knot ends stay fixed
for at least 20 000 000 $\tau$ - a duration which is at least
three orders of magnitude longer that optimal folding times
of unknotted proteins within the same model \cite{biophysical}.
At $T=1.00~ \epsilon/k_B$, the knot ends stay put for a while and
then diffuse out of the chain rapidly. At $t=0.95~\epsilon/k_B$ we
observe an intermittent behavior in which the knot disappears and is
then restored. If there is any intermittency at $T=0.9~\epsilon/k_B$,
then the recovery time is longer than the scale of the simulations.
The bottom panel of Fig. \ref{probab} shows the median unfolding time
defined as in ref. \cite{thermal}, {\it i.e.} through the instant
at which all contact that are sequentially separated by more than $l$
are broken simultaneously. Taking $l$ of 4 (local helical contacts)
was numerically unfeasible so we took $l=10$. The statistics were based
on 110 trajectories. At $T=1.0~\epsilon/k_B$ and below less than 50\%
trajectories led to folding and the median time could not be determined
-- with the cutoff of 30 000 $\tau$.

\subsection{Folding in unbounded space}

We now consider folding in an unbounded space when one starts from a 
random conformation which is nearly fully extended. 
Even though we use the same model as in ref. \cite{dodging},
we do not find even one single trajectory 
that would lead to correct folding. However, there were trajectories
which led to the establishment of all native contacts in an unknotted way.
We shall refer to such situations as corresponding to misfolding.
One of them is shown in panel C of Fig. \ref{fig:folded}.
The lack of proper folding in 1J85 essentially  agrees with
the result of ref. \cite{Takada} (where 0.1\% success rate is reported). 
We have generated 1201 trajectories at
$0.35 ~\epsilon/k_B$ and between 200 and 314 trajectories at 0.1, 0.2 and 
$0.3~ \epsilon/k_B$. The simulational cutoff time was 1 000 000 $\tau$.
With this statistics, a 1 - 2\% success rate claimed in ref. \cite{dodging}
(at unspecified temperatures)
would mean observation of good folding in at least 12 trajectories.
We speculate that the discrepancy may be due to { the following factors
a)} possible biases in the initial 
conformations used in that reference,
{ b)} malfunctioning of the KMT
algorithm (that may show knots when none are present), {
and c) the folding temperature range is narrower than 0.025 $\epsilon/k_B$,
i.e. the steps we considered, and the right $T$ to fold was found
accidentally.}
It should be noted that we have 
obtained a total of 338 misfolded trajectories (17\% of all trajectories).
143 of these are
at $0.35~\epsilon/k_B$ and correspond to a mean folding time of 472 651 $\tau$.
{ It is possible to mistake some of the misfolded states for 
the knotted ones.}
{ We find no folding nor misfolding at $T_f$ defined by the
condition of $Q=\frac{1}{2}$.}

Not much is changed when one attempts to reduce the conformational entropy
by anchoring the C terminus. For the total of 427 trajectories (with
$T$ between 0.1 and $0.4~\epsilon/k_B$) 93 ({\it i.e.} 22\%) established 
all native contacts without forming the knot. 

We have also considered a different Go-like model, along the lines
of Clementi {\it et al.} \cite{Clementi}. In this model, the backbone
stiffness is accounted for by the more common bond and dihedral angle
potentials. The contact interactions are given by the 10-12 Lennard-Jones
potentials, and we take the contact map to be given by OV 
(here we we removed the $i,i+2$ and $i,i+3$ contacts). For this model,
$T_f$ moves upward by about 0.2 $\epsilon/k_B$ and $P_0$ is
about 0.01 at $0.6~\epsilon/k_B$. At this $T$, only one out of 246 
trajectories led to the correct folding with the proper knot
(the folding time was 510 264 $\tau$). 
At temperatures 0.7$~\epsilon/k_B$ and 0.75$~\epsilon/k_B$ we 
get 1\% and 7\% of the correctly folded trajectories respectively and 
none at lower temperatures. This model, however, is not the one
considered in ref. \cite{dodging}.

\subsection{Folding on the ribosome}

\begin{figure}[ht]
\begin{center}
\includegraphics[scale=0.8]{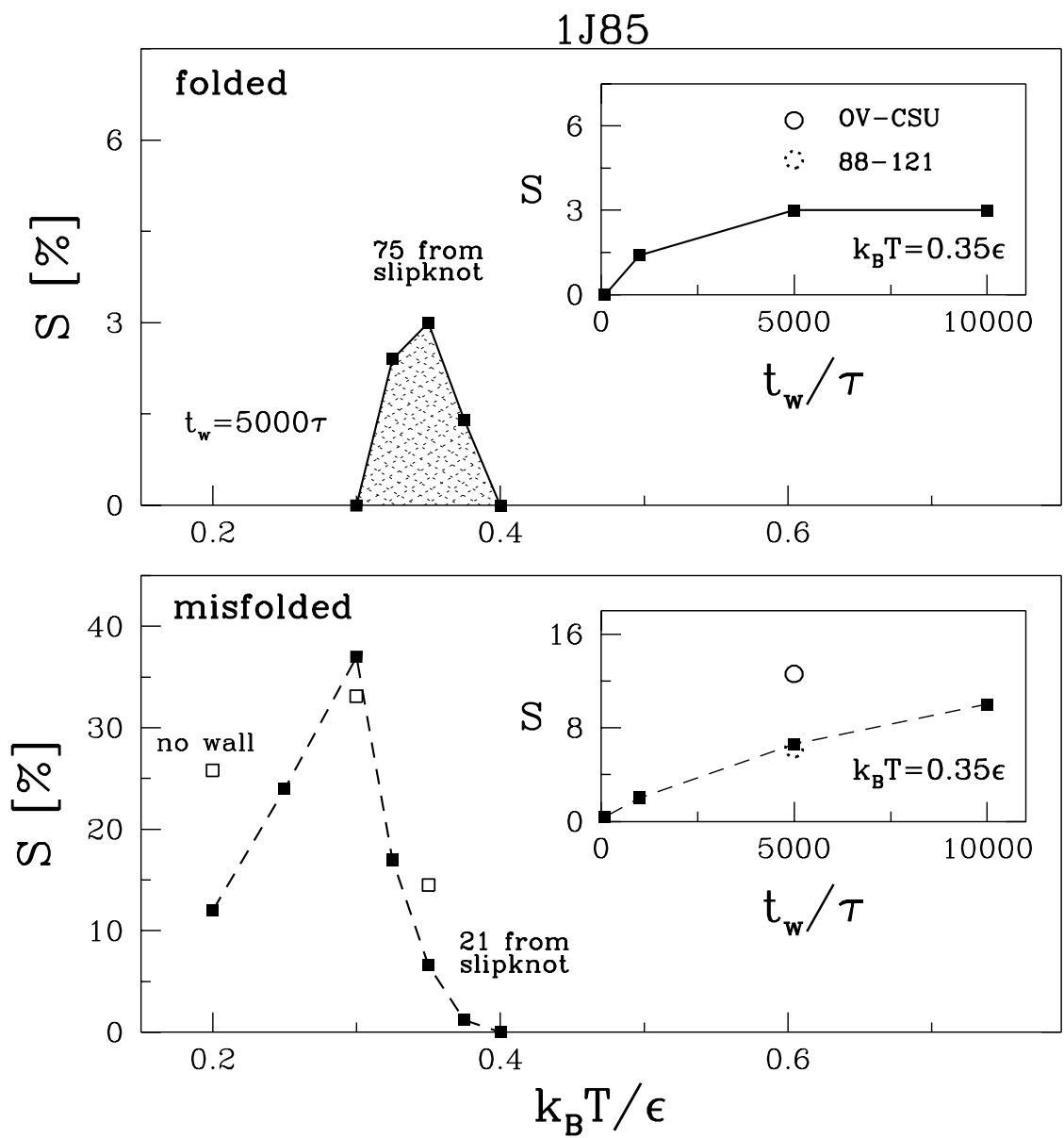}
\end{center}
  \caption{1J85: the kinetic results for the on-ribosome folding.
The upper panel shows the percentage-wise success in folding as
a function of $T$ for $t_w$=5000$~\tau$. The inset shows $S$
as a function of $t_w$ for the optimal temperature of $0.35~\epsilon/k_B$.
The dotted circle shows $S$ when one contact, 88-121 is added.
The full circle shows $S$ when all specific CSU contacts are
added. The lower panel with its inset is similar but shows the
data for misfolding -- when the native contacts get established
but the knot is not formed.
}
  \label{knotfol}
\end{figure}

The percentage-wise success, $S$, of reaching the properly knotted
folded conformation increases substantially by simulating
the process in the cotranslational way. The results are summarized
in Fig. \ref{knotfol}. They depend on the value of $t_w$ but there 
appears to be a saturation in $S$ around $t_w$ of 5 000$~\tau$.
The experimental times of translation are certainly much
longer than $\sim$ 5~$\mu$s but this is not expected to affect the $S$.
However, the corresponding $S$ for misfolding (the inset in the 
lower panel) may reach saturation at a bigger $t_w$.

The data for the evidently optimal $0.35 \epsilon/k_B$
were obtained based on 500 trajectories for each $t_w$ that was
considered. Same statistics were used for 0.325 and 0.375$~\epsilon/k_B$
and $t_w$ of 5 000$~\tau$. In other cases, there were at least 100 
trajectories. The average combined time of folding under the
optimal conditions with $t_w=5000~\tau$ is 860353$~\tau$. Comparable
times are at 0.325 and $0.375~\epsilon/k_B$. 

Interestingly, switching
to the Clementi {\it et al.}-like model generates no proper folding
independent of whether the contact map is OV, CSU, or OV-CSU
(in the temperature range between 0.45 and $1.3~\epsilon/k_B$).
The backbone chain appears to be too stiff to allow for a knot
in this model.

The lower panel of Fig. \ref{knotfol} shows that 
the time evolution from extended states results in a substantial
percentage of the misfolded conformations. This percentage gets boosted 
by the nascent conditions from 33\% to 37\% if the evolution takes
place at $0.3\epsilon/k_B$. At this $T$, there is no knotted folding.
On the other hand, at the knot-optimal temperature of $0.35\epsilon/k_B$
the nascent conditions produce only 6.6\% of the misfolded states.

\begin{figure}[ht]
\begin{center}
\includegraphics[scale=0.9]{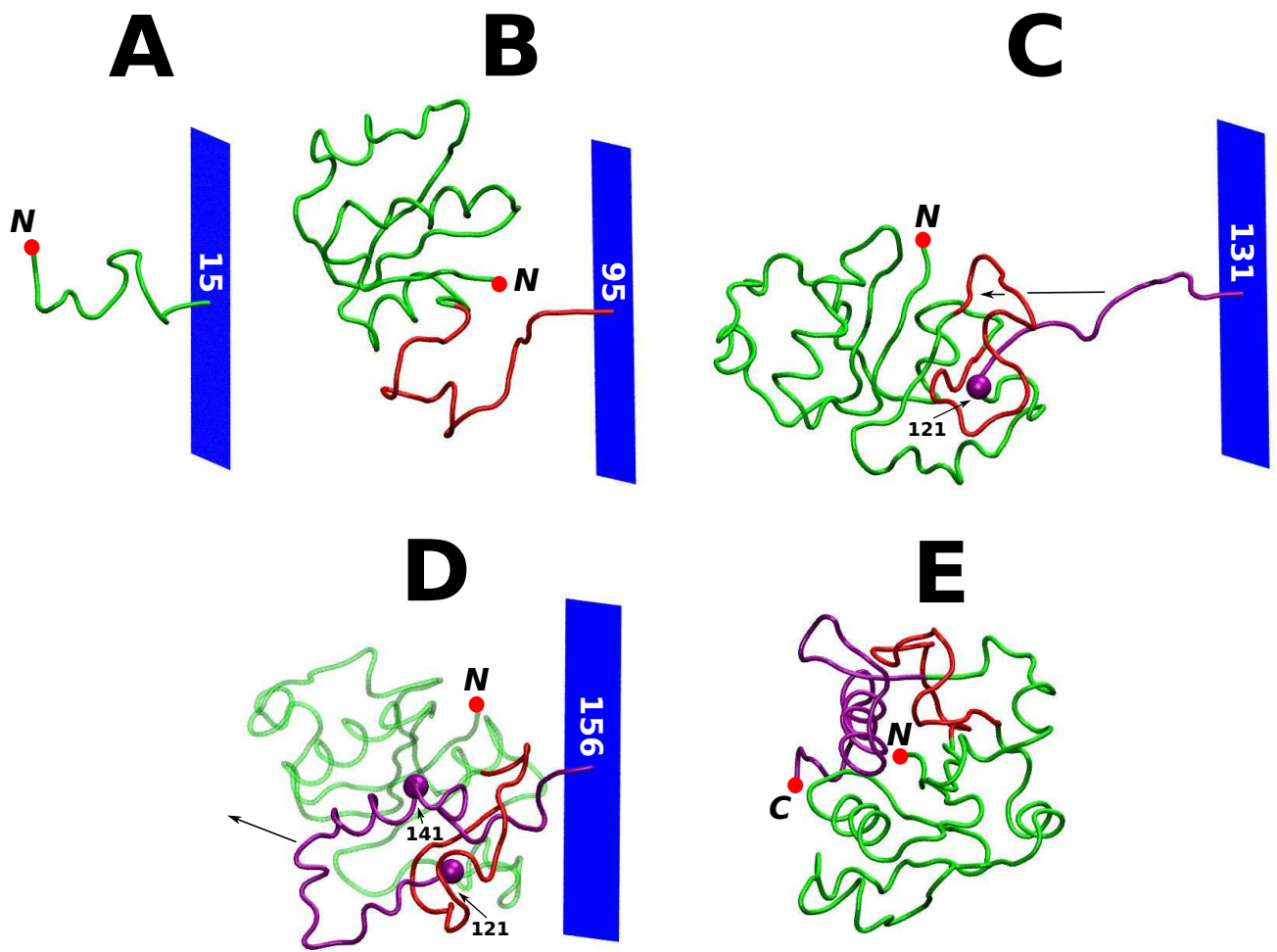}
\end{center}
  \caption{ 
Stages of folding and knotting process of protein 1J85 in the presence 
of the repulsive wall. The wall is represented by the blue plate. The 
number written on the plate indicates the residue that has just been born.
The part of the protein which creates the loop is marked in red color. 
Residue 121 is represented by a purple ball in panel C -- it is the
begining of a slipknot part. 
The arrows in panels C and D indicate the direction in which the
remaining part of the backbone moves through the red loop. 
The completely knotted, folded and released protein is presented 
in panel E without the wall as the protein gets detached.
}
  \label{knotwall}
\end{figure}

\begin{figure}[ht]
\begin{center}
\includegraphics[scale=1.2]{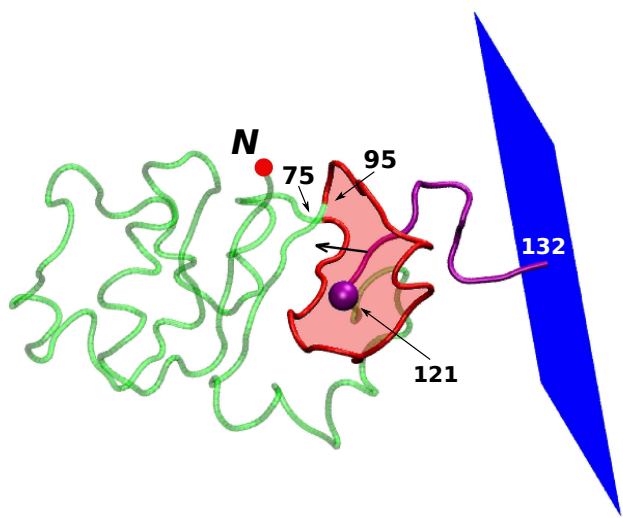}
\end{center}
  \caption{ 
Conformation of the nascent 1J85 just before formation of the slipknot
state. Residue 121, represented by a purple ball,
stays pinned to the shaded surface spanned by the red knot-loop.
In order to form the knot, the remaining C-terminal part of the
protein has to drag through the knot-loop in the direction shown by the
arrow.
}
  \label{loopy}
\end{figure}

\begin{figure}[ht]
\begin{center}
\includegraphics[scale=0.6]{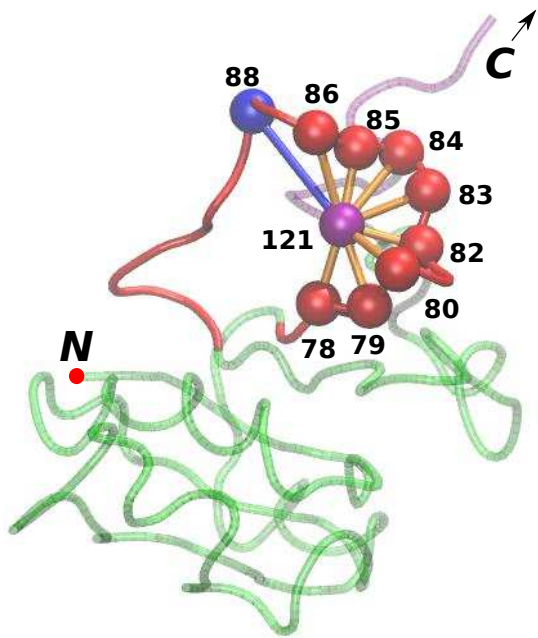}
\end{center}
  \caption{The contacts involving residue 121 in 1J85. The contacts 
shown in the orange color are derived through the overlap criterion.
The contact shown in the blue color is an extra contact derived
through the CSU approach. All of these contacts are with the residues
that are a part of the red knot-loop. 
}
  \label{contacts}
\end{figure}

\begin{figure}[ht]
\begin{center}
\includegraphics[scale=0.6]{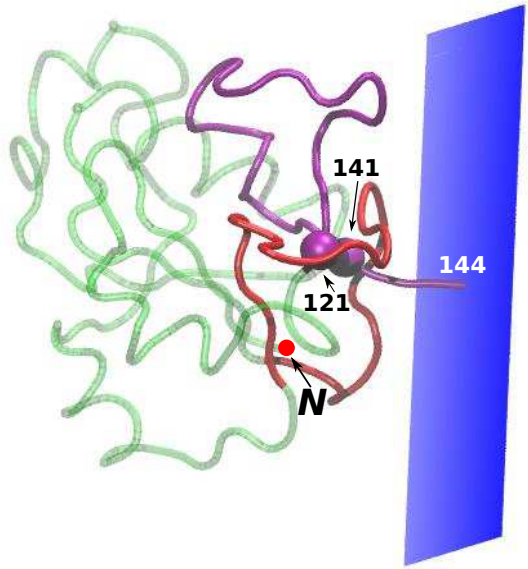}
\end{center}
  \caption{The fully formed slipknot state in 1J85: the purple part of the
backbone just went through the red knot-loop. The purple spheres indicate
the begining (121) and ending (141) residues in the slip-loop. }
  \label{slipknot}
\end{figure}

Fig. \ref{knotwall} shows five snapshots of an example of a successful
folding trajectory in our basic model of 1J85. 
(A related movie is available as the Supplementary Material).
The sequential fragments
emerge from the model ribosome in order: first green, then red, green again,
and finally purple. At stage A, there is no tertiary order yet.
At stage B, the knot-loop segment (75-95) has left the model ribosome.
At stage C, the knot-loop forms a nearly planar and nearly closed
contour and residue 121 arrives at the plane of this contour. 
Another perspective on this stage is shown in Fig. \ref{loopy}
where the knot-loop is shaded in red. The C-terminal
part (121-156, in purple) of the protein must drag through the knot-loop
and the success rate depends on how well residue 121 is pinned 
to the plane of the knot-loop. There are eight OV-based contacts that
residue 121 makes with the knot-loop, as shown in Fig. \ref{contacts}.
However, the stabilization of the attachment is enhanced by the
CSU-derived contact 88-121. A fully formed slipknot is shown in panel
D of Fig. \ref{knotwall} and, in  more details, in Fig. \ref{slipknot}.
At the very next stage E, the protein gets detached from the ribosome and 
becomes knotted because the C-terminal segment goes through the
knot-loop.

Our results confirm the picture proposed in ref. \cite{dodging}
that knotting in 1J85 is enabled by the slipknotting mechanism.
However, we see it operational only when the protein is nascent.
The wall facilitates formation of the C-terminal slipknot
on the correct side of the knot-loop and it provides
semi-confinement that allows for making repeated attempts
to drag the slip-loop through the knot-loop.
When one removes the wall but starts the evolution from the
slipknot conformation, then -- at optimality -- 75\% of trajectories
lead to the knot formation.

It is interesting to note that inclusion of the 88-121 contact, 
at $T=0.35\epsilon/k_B$ and for $t_w$=5000$~\tau$,
boosts $S$ from 3.0\% to 4.8\%  An inclusion of all missing CSU
contacts (the OV-CSU contact map) boosts it even further to 6.2\%. 
All of these contacts should be considered native.
Wallin, {\it et. al.} \cite{Wallin} have argued that non-native 
contacts are necessary to fold 1J85 to the knotted state.
Specifically they added exponentially decaying non-native
contacts between segments [86,93] and [122,147].
Their definition of a native contact is that two heavy atoms
from different residues must fall within the distance of 4.5 {\AA}
from one another. This procedure misses the 88-121 contact.
In our contact map, we generate 13 OV native contacts for the segments
chosen by Wallin {\it et al.}. CSU adds 4 more and 88-121 is the
fifth one. Thus there is no need to invoke non-native contacts
to explain folding in 1J85.

\subsection{The energetics of the slipknot formation during folding on the ribosome}

\begin{figure}[ht]
\begin{center}
\includegraphics[scale=1.2]{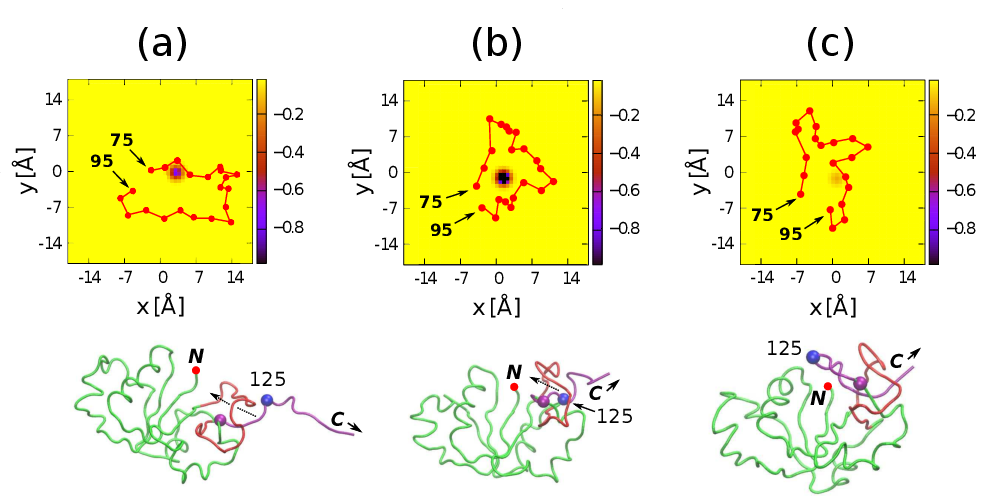}
\end{center}
  \caption{ 
Three successive stages of the successful knotting process of protein 1J85
in the presence of the repulsive wall. The upper panels
show the potential energies associated with amino acid 125
in units of $\epsilon$, as indicated by the color scales on the right.
The chain shows the projection of the knot-loop onto the $x-y$ plane which is
parallel to the repulsive plane.  The origin (0,0) is set at location of 
amino acid 125. The lower panels show the corresponding conformations of the
protein. The thick arrow next to C indicate the sequential direction
toward the C terminus which is not yet born.
The dotted arrows indicate the direction of movement of amino acid 125.
Panel (a) is for the situation in which the slipknot is about to be formed
and panel (c) just after it was fully formed.
}
  \label{energiesgr}
\end{figure}

We have observed that the knotting process is very rapid. It takes
quite a long time for a protein to get to state (a) shown in Fig. \ref{energiesgr},
in which the slip-loop is positioned just above the knot-loop.
We expect that 
the specific amino-acid arrangement
leads to formation of a potential well and, therefore, emergence
of a force that drags the slip-loop through the knot-loop.
To prove this, we have monitored the potential energy
associated with specific amino acids from the slip-loop.
The top panels in Fig. \ref{energiesgr}  show the energy
experienced by one of the amino acids from the slip-loop, 
the 125th in the sequence,
if found at various locations
within the plane parallel to  the plane of the model
ribosome and crossing through this amino acid. 
The potential well is very localized. In state (a), its depth is
close to $0.50~\epsilon$  and in state (b), 200 $\tau$ later,
it increases to $1.0~\epsilon$.
In state (c), the slipknot is already created and
the well becomes very shallow -- 0.07~$\epsilon$ -- and the slipknot
ceases to move forward any further.
If at stage (a) the conformation of the knot-loop is such that
the well is not sufficiently deep then no proper knotting takes place.

\subsection{Other {deeply} knotted proteins}

We have considered two other {deeply} knotted (hypothetical)
proteins \cite{1O6D}: 1O6D 
from {\it Thermatoga maritima} and 1VH0 from {\it Staphylococcus aureus}.
Both were analyzed previously through stretching simulations \cite{Sulkowska_2008}.
We find no properly folded trajectories for 1O6D and 1VH0, either with
or without the wall. However, we have obtained substantial percentages
of the misfolding situations. For 1O6D, 41 out of 400 trajectories
without the wall and 151 out of 400 with the wall resulted in misfolding.
For 1VH0, the corresponding numbers are 136 out of 500 and 203 out of 390.
Thus the ribosome-imitating wall helps in folding through semi-confinement
but not sufficiently enough to form the proper knots in a noticeable way.
However, more sophisticated models of the ribosome might work better.

\section{Conclusions}

We have used the structure-based molecular dynamics model to study
folding of {deeply} knotted proteins. The structure-based models favor
formation of the native contacts but carry no topological bias.
It is an open question how to implement such a bias in a model.
Establishment of the native contacts usually does not mean
establishing the knot. The nascent conditions 
are found to enhance the probability 
of establishing the contacts and sometimes 
also the formation of the knot.
We also find that achieving proper folding requires staying within the
proper range of temperatures.

Our results are consistent with the experiments of Mallam {\it et al.}
\cite{Mallam} that nascent proteins form knots more easily
and that knots in 1J85 (YibK) persist in the chemically denatured state \cite{Mallam1}.
Independent of this, the GroEL-GroES chaperonin complex may also accelerate
knotted folding \cite{Mallam}.

Our results for 1J85 support the slipknot mechanism in the knot
formation but suggest that it may operate only under the nascent 
conditions and when the temperature is within an optimal range. 
{The geometry of the ribosome tunnel should provide even more
confinement than that given just by the plane and should boost the
efficiency of cotranslational folding.}
Finally, there is no need to invoke non-native
contacts to fold to the knotted state in this system.

Proteins with shallow knots, such as 
MJ0366 from {\it Methanocaldococcus jannaschii}
with the PDB structure code of 2EFV, studied
theoretically in refs. \cite{Micheletti,Noel} appear to
fold in a different way both off- and on-ribosome. A discussion
of this problem is being prepared for a separate report.

{\bf Acknowledgments}
We appreciate useful discussions with J. Trylska.
This work has been supported by 
by the ERA-NET-IB/06/2013 Grant FiberFuel funded by the National Centre for 
Research and Development in Poland.
The local computer resources were financed by the European Regional Development Fund 
under the Operational Programme Innovative Economy NanoFun POIG.02.02.00-00-025/09.

\clearpage

\begin{thebibliography}{100}

\bibitem{Virnau}
Virnau P, Mirny L A, Kardar M 2006
\textit{Plos. Comp. Biol.} \textbf{2} e122

\bibitem{Virnau1}
Virnau P, Mallam A, Jackson S 2011
\textit{J. Phys. Cond. Mat.} \textbf{23} 033101

\bibitem{Stasiak}
Su{\l}kowska J I, Rawdon E J Millet K C, Onuchic J N, Stasiak A 2012
\textit{Proc. Natl. Acad. Sci. USA} \textbf{109} E1715-E1723

\bibitem{km1}
Koniaris K, Muthukumar M 1991
\textit{Phys. Rev. Lett.} \textbf{66} 2211-2214

\bibitem{taylor}
Taylor W 2000
\textit{Nature} \textbf{406} 916-919

\bibitem{Sulkowska_2008}
Su{\l}kowska J I, Su{\l}kowski P, Szymczak P, Cieplak M 2008
\textit{Phys. Rev. Lett.} \textbf{100} 058106

\bibitem{JACS}
Su{\l}kowska J I, Su{\l}kowski P, Szymczak P, Cieplak M 2010
\textit{J. Am. Chem. Soc.} \textbf{132} 13954-13956

\bibitem{Dziubiella}
Dziubiella J 2009
\textit{Biophys. J.} \textbf{96} 831-839

\bibitem{Israel}
Chwastyk M, Cieplak M 2014
\textit{Israel Journal of Chemistry} \textbf{54} 1241-1249

\bibitem{Dill}
Dill K A, MacCallum J L 2012
\textit{Science} \textbf{338} 1042-1046

\bibitem{dodging}
Su{\l}kowska J I, Su{\l}kowski P, Onuchic J N 2009
\textit{Proc. Natl. Acad. Sci. USA} \textbf{106} 3119-3124

\bibitem{Takada}
Li W, Terakawa T, Wang W, Takada S 2012
\textit{Proc. Natl. Acad. Sci. USA} \textbf{109} 17789-17794

\bibitem{Dobson}
Cabrita L D, Dobson C M, Christodoulou J 2010
\textit{Curr. Op. Struct. Biol.} \textbf{20} 33-45

\bibitem{Bustamante}
Kaiser C M, Goldman D H, Chodera J D, Tinoco Jr. I, Bustamante C 2011
\textit{Science} \textbf{334} 1723-1727

\bibitem{ribosome}
Melnikov S, Ben-Shem A, Garreau de Loubresse N, Jenner L, Yusupova G, Yusupov M 2012
\textit{Nat. Struct. Mol. Biol.} \textbf{19} 560-567

\bibitem{ribosome1}
Garreau de Loubresse N, Prokhorova I, Holtkamp W, Rodnina M V,
Yusupova G, Yusupov M 2014
\textit{Nature} \textbf{513} 517-522

{
\bibitem{ribosome2}
Voss N R, Gerstein M, Steitz T A, Moore P B 2006
\textit{J. Mol. Biol.} \textbf{360} 893-906

\bibitem{ribosome3}
Frank J, Gonzales R L Jr. 2010
\textit{Annu. Rev. Biochem.} \textbf{79} 381-412
}

\bibitem{1O6D}
Badger J, Sauder J M, Adams J M {\it et al.} 2005
\textit{Proteins} \textbf{60} 787-796

\bibitem{1J85}
Lim K, Zhang H, Tempczyk A, Krajewski W, Bonander N, Toedt J, Howard A, Eisenstein E, Herzberg O 2003 
\textit{Proteins} \textbf{51} 56-67

\bibitem{Go0}
Go N 1983 {\textit Annu. Rev. Biophys. Bioeng.} \textbf{12} 183-210

\bibitem{JPCM}
Su{\l}kowska J I and Cieplak M 2007
\textit{J. Phys.: Cond. Mat.} \textbf{19} 283201

\bibitem{models}
Su{\l}kowska J I, Cieplak M 2008 
{\textit Biophys. J.} \textbf{95} 3174-3191

\bibitem{PLOS} Sikora M, Su{\l}kowska J I, Cieplak M 2009
\textit{PLOS Comp. Biol.} \textbf{5} e1000547

\bibitem{Tsai}
Tsai J, Taylor R, Chothia C, Gerstein M 1999 
\textit{J. Mol. Biol.} \textbf{290} 253-266

\bibitem{Sobolev}
Sobolev V, Sorokine A, Prilusky J, Abola E E, Edelman M 1999
\textit{Bioinformatics} \textbf{15} 327-332

\bibitem{Elcock}
Elcock A H 2006
\textit{PLOS Comp. Biol.} \textbf{2} e98

\bibitem{BSDB}
Sikora M, Su{\l}kowska J I, Witkowski B S, Cieplak M 2010
\textit{Nucl. Acids Res.} \textbf{39} D443-D450

\bibitem{Socci}
Socci N D, Onuchci J N 1994
\textit{J. Chem. Phys.} \textbf{101} 1519-1528

\bibitem{Banavar}
Cieplak M and Banavar J R 2013
\textit{Phys. Rev. E. Rapid Comm.} \textbf{ 88}, 040702(R)

\bibitem{Banavar1}
Cieplak M and Banavar J R 2013
\textit{EPL} \textbf{104} 58001

\bibitem{biophysical}
Cieplak M, Hoang T X 2003 
\textit{Biophys. J.} \textbf{84} 475-488

\bibitem{thermal}
Cieplak M, Su{\l}kowska J I 2005
\textit{J. Chem. Phys.} \textbf{123} 194908

\bibitem{Clementi}
Clementi C, Nymeyer H, Onuchic J N 2000 
\textit{J. Mol. Biol.} \textbf{298} 937-953

\bibitem{Wallin}
Wallin S, Zeldovich K B, Shakhnovich E I 2007
\textit{J. Mol. Biol.} \textbf{368} 884-893

\bibitem{Mallam}
Mallam A L, Jackson S E 2012
\textit{Nat. Chem. Biol.} \textbf{8} 147-153

\bibitem{Mallam1}
Mallam A L, Rogers J M, Jackson S E 2010
\textit{Proc. Natl. Acad. Sci. USA} \textbf{107} 8189-8194

\bibitem{Micheletti}
Beccara S A, Skrbic T, Covino R, Micheletti C, Faccioli P 2013
\textit{PLOS Comp. Biol.} \textbf{9} e1003002

\bibitem{Noel}
Noel J K, Onuchic J N, Su{\l}kowska J I 2013
\textit{Phys. Chem. Lett.} \textbf{4} 3570-3573



\end{thebibliography}
\end{document}